\documentclass[twocolumn,superscriptaddress,nofootinbib,aps,prl]{revtex4-2}

\usepackage{graphicx}
\usepackage{dcolumn}
\usepackage{bm}
\usepackage{amsmath}
\usepackage{amssymb}
\usepackage{siunitx}
\usepackage{upgreek}
\usepackage{hyperref}
\usepackage{overpic}
\usepackage[normalem]{ulem}

\begin{document}

\title{A New Bound on Light-by-Light Scattering in the Low-Energy Regime}

\newcommand{\HZDR}{\affiliation{Helmholtz-Zentrum Dresden-Rossendorf, Bautzner Landstra\ss e 400, 01328 Dresden, Germany}}
\newcommand{\HIJ}{\affiliation{Helmholtz-Institut Jena, Fr\"obelstieg 3, 07743 Jena, Germany}}
\newcommand{\GSI}{\affiliation{GSI Helmholtzzentrum f\"ur Schwerionenforschung, Planckstra\ss e 1, 64291 Darmstadt, Germany}}
\newcommand{\FSU}{\affiliation{Physikalisch-Astronomische Fakultät, Friedrich-Schiller-Universität Jena, 07743 Jena, Germany}}
\newcommand{\XFEL}{\affiliation{European XFEL, Holzkoppel 4, 22869 Schenefeld, Germany}}
\newcommand{\IOP}{\affiliation{Department of Radiation and Chemical Physics, Institute of Physics, Czech Academy of Sciences, Na Slovance 2, 182 00 Prague 8, Czech Republic}}
\newcommand{\CUNI}{\affiliation{Department of Surface and Plasma Science, Faculty of Mathematics and Physics, Charles University, V Hole\v{s}ovi\v{c}k\'ach 2, 182 00 Prague 8, Czech Republic}}
\newcommand{\CTU}{\affiliation{Department of Nuclear Chemistry, Faculty of Nuclear Science and Physical Engineering, Czech Technical University in Prague, B\v{r}ehov\'a 7, 115 19 Prague 1, Czech Republic}}
\newcommand{\IPP}{\affiliation{Laser Plasma Department, Institute of Plasma Physics, Czech Academy of Sciences, U Slovanky 2525/1a, 182 00 Prague 8, Czech Republic}}
\newcommand{\DESY}{\affiliation{Deutsches Elektronen-Synchrotron DESY, Notkestra\ss e 85, 22607 Hamburg, Germany}}
\newcommand{\TUD}{\affiliation{Technische Universit\"at Dresden, 01062 Dresden, Germany}}
\newcommand{\CASUS}{\affiliation{Center for Advanced Systems Understanding, Untermarkt 20, 02826 G\"orlitz, Germany}}
\newcommand{\PLY}{\affiliation{Mathematical Sciences Department, University of Plymouth, Plymouth, PL4 8AA, United Kingdom}}
\newcommand{\ELI}{\affiliation{ELI Beamlines Facility, The Extreme Light Infrastructure ERIC, Doln\'i B\v{r}e\v{z}any, Czech Republic}}
\newcommand{\ESRF}{\affiliation{European Synchrotron Radiation Facility, CS 40220, F-38043 Grenoble Cedex 9, France}}
\newcommand{\PALM}{\affiliation{PALM Scientific LLC, Naperville, IL, USA}}

\author{Aim\'e~Matheron$^{\,\parallel}$}\email{a.matheron@hi-jena.gsi.de}\HIJ\GSI\FSU
\author{Michal~\v{S}m\'id$^{\,\parallel}$}\email{m.smid@hzdr.de}\HZDR

\author{Felix Karbstein}\email{f.karbstein@hi-jena.gsi.de}\HIJ\GSI\FSU
\author{Sebastian~G\"ode}\XFEL
\author{Willi~Hippler}\HIJ\GSI\FSU
\author{Robert~L\"otzsch}\HIJ\GSI\FSU

\author{Naser~Ahmadiniaz}\HZDR
\author{Michal~Andrzejewski}\XFEL
\author{Sergey~Antipov}\PALM
\author{Carsten~Baehtz}\HZDR
\author{Christina~B\"omer}\DESY
\author{Victorien~Bouffetier}\HZDR
\author{Erik~Brambrink}\XFEL
\author{Jakub~Buli\v{c}ka}\IOP\CUNI
\author{Samuele~Di~Dio~Cafiso}\HZDR
\author{Rafael~Celestre}\ESRF
\author{Jarom\'ir~Chalupsk\'y}\IOP
\author{Thomas~E.~Cowan}\HZDR\TUD
\author{James~P.~Edwards}\PLY
\author{Sebasti\'an Franchino-Vi\~nas}\altaffiliation{Present address: Departamento de Física, Facultad de Ciencias Exactas, Universidad Nacional de La Plata, C.C. 67 (1900), La Plata, Argentina}\HZDR
\author{Holger~Gies}\FSU\HIJ
\author{J\"org~Grenzer}\HZDR
\author{Travis Griffin}\XFEL
\author{V\v{e}ra~H\'ajkov\'a}\IOP
\author{Jens~Hauser}\HZDR
\author{Thomas~Heinzl}\PLY
\author{Uwe~Hernandez~Acosta}\HZDR\CASUS
\author{Peter~Hilz}\HIJ\GSI\FSU
\author{Hauke~H\"oppner}\HZDR
\author{Al\v{z}b\v{e}ta~Horynov\'a}\IOP\CTU
\author{Lingen~Huang}\HZDR
\author{Oliver~Humphries}\XFEL
\author{\v{S}imon~Jel\'inek}\IOP\IPP\CUNI
\author{Libor~Juha}\IOP
\author{Pooyan~Khademi}\HIJ\GSI\FSU
\author{Ben~King}\PLY
\author{Christian~Kohlf\"urst}\HZDR
\author{Dietrich~Krebs}\DESY
\author{Alejandro~Laso-Garc\'ia}\HZDR
\author{Alexander~J.~Macleod}\ELI
\author{Berit~Marx-Glowna}\HIJ\GSI
\author{Masruri~Masruri}\HZDR
\author{Nikita~Medvedev}\IOP
\author{Anthony~Mercuri-Baron}\PLY
\author{Elena~A.~Mosman}\FSU\HIJ
\author{Motoaki~Nakatsutsumi}\XFEL
\author{Gerhard~G.~Paulus}\HIJ\GSI\FSU
\author{Jennifer~Popp}\XFEL
\author{Thomas~R.~Preston}\XFEL
\author{Isabella~Probst}\HIJ\GSI\FSU 
\author{Sripati~V.~Rahul}\XFEL
\author{Lisa~Randolph}\XFEL
\author{Ralf~R\"ohlsberger}\DESY\HIJ\GSI\FSU
\author{Thomas~Roth}\ESRF
\author{Sakshath~Sadashivaiah}\HIJ
\author{Felipe~C.~Salgado}\HIJ\GSI\FSU 
\author{Roland~Sauerbrey}\HZDR\TUD
\author{Alexander~S\"avert}\XFEL
\author{Hans-Peter~Schlenvoigt}\HZDR
\author{Sebastian~M.~Schmidt}\HZDR\TUD
\author{Ulrich~Schramm}\HZDR\TUD
\author{Ralf~Sch\"utzhold}\HZDR\TUD
\author{Jan~Patrick~Schwinkendorf}\HZDR\XFEL
\author{Frank~Seiboth}\DESY
\author{Daniel~Seipt}\HIJ\GSI\FSU
\author{Georgiy~Shoulga}\HZDR
\author{Thomas~St\"ohlker}\HIJ\GSI\FSU
\author{Minxue~Tang}\XFEL
\author{Abha~Thazhathe~Peedika}\HIJ\GSI\FSU
\author{Monika~Toncian}\HZDR
\author{Toma~Toncian}\HZDR
\author{Maksim~Valialshchikov}\HIJ\GSI\FSU
\author{Vojt\v{e}ch~Vozda}\IOP
\author{Edgar~Weckert}\DESY
\author{Colin~Wessel}\HIJ\GSI\FSU
\author{Jan~Wild}\CUNI
\author{Andreas~Wipf}\FSU
\author{Ulf~Zastrau}\XFEL

\author{Matt~Zepf}\email{m.zepf@hi-jena.gsi.de}\HIJ\GSI\FSU

\collaboration{BIREF@HIBEF Collaboration}\noaffiliation

\date{\today}

\begin{abstract}
The superposition principle at the heart of optics states that electromagnetic fields do not interact in vacuum. This principle is not universal, instead electromagnetic fields couple weakly via quantum vacuum fluctuations. Striking consequences are light-by-light scattering and vacuum birefringence.
Experimental results from colliding X-ray free-electron and optical high-intensity lasers combined with dark-field detection and polarimetry for background suppression are reported. This approach yields the most stringent bound for light-by-light scattering in the low-energy regime achieved to date -- improving upon previous experimental limits by ten orders of magnitude -- and paves the way for precision measurements.
\end{abstract}

\maketitle

\begingroup
\renewcommand\thefootnote{\(\parallel\)}
\footnotetext[1]{These authors contributed equally to this work.}
\endgroup

\paragraph{Introduction}
At the foundations of optics lies the superposition principle of electromagnetic waves in vacuum implying that any two light waves traverse each other without interaction.
This is a direct consequence of the equations governing the electromagnetic field being linear, which aligns perfectly with our everyday experience.
However, since the seminal works of Euler, Heisenberg, and Kockel dating back to the 1930s \cite{Euler:1935zz,Euler:1935qgl,Heisenberg:1936nmg}, it has been known that this  principle is not universal, but will ultimately be modified by quantum corrections associated with the vacuum fluctuations of charged particle fields.
Within the Standard Model (SM) of particle physics, the leading correction arises from quantum electrodynamics (QED) and stems from the effective coupling of electromagnetic fields mediated by virtual electron-positron pairs.
This results in nonlinear equations governing the electromagnetic field and gives rise to  observable consequences such as light-by-light scattering and vacuum birefringence, both of which have so far eluded  experimental verification in the low-energy regime, characterized by photon energies well below the electron mass \cite{Ahmadiniaz:2024xob}.

To be specific, in the low-energy, weak-field regime the equations governing the electromagnetic field become ($c=\hbar=1$)
\begin{multline}
\partial_\mu F^{\mu\nu}\,\simeq\,\frac{1}{45}\frac{\alpha}{\pi}\Bigl(\frac{e}{m^2}\Bigr)^2 \partial_\mu
\Bigl(a\,{\cal F}\,
F^{\mu\nu}+
b\,{\cal G}\,
{}^\star\!F^{\mu\nu}\Big)\,, \label{eq:EoM}
\end{multline}
with ${\cal F}=F_{\mu\nu}F^{\mu\nu}/4$ and ${\cal G}=F_{\mu\nu}{}^\star\!F^{\mu\nu}/4$ determined by contractions of the field strength tensor $F^{\mu\nu}$ and its dual ${}^\star\!F^{\mu\nu}$;  $\alpha$ is the fine-structure constant, $e$ is the elementary charge, and $m$ is the electron mass.
Note, that the right-hand side of Eq.~\eqref{eq:EoM} is cubic in the electromagnetic field and can be considered as the source of   the quantum vacuum signal \cite{McKenna:1963}.
The dimensionless coefficients $a$ and $b$ parameterize the strength of the two distinct nonlinear self-coupling terms of the electromagnetic field.
QED predicts these to be given by \cite{Heisenberg:1936nmg,Ritus:1975pcc,Dittrich:1985yb,Gies:2016yaa}
\begin{equation}
a=4\Bigl(1+\frac{40}{9}\frac{\alpha}{\pi}+\ldots\Bigr)\ \ \ \text{and}\ \ \
b=7\Bigl(1+\frac{1315}{252}\frac{\alpha}{\pi}+\ldots\Bigr)\,. \label{eq:a,b}
\end{equation}
In the classical limit ($a=b=0$), corresponding to the absence of vacuum fluctuations, the right-hand side of Eq.~\eqref{eq:EoM} vanishes and linear Maxwell theory in vacuum is recovered. Experimentally, however, direct observation of optical activity of the vacuum is challenging. Numerous experiments have been conducted to date aiming to set tighter and tighter bounds with the ultimate aim of a positive measurement of quantum vacuum nonlinearities.  Experiments using colliding photon beams in the low energy regime are of particular relevance to the results reported here \cite{Moulin:1996vv,Bernard:2000ovj,Yamaji:2016xws,Watt:2024brh}. The Polarizzazione del Vuoto con LASer (PVLAS) experiment~\cite{Ejlli:2020yhk} which aimed to measure vacuum birefringence induced by a magnetic field and probed by a photon beam is particularly noteworthy for  establishing the most stringent laboratory bound on vacuum birefringence to date.

\begin{figure*}
	\centering
	\includegraphics[width=1\linewidth]{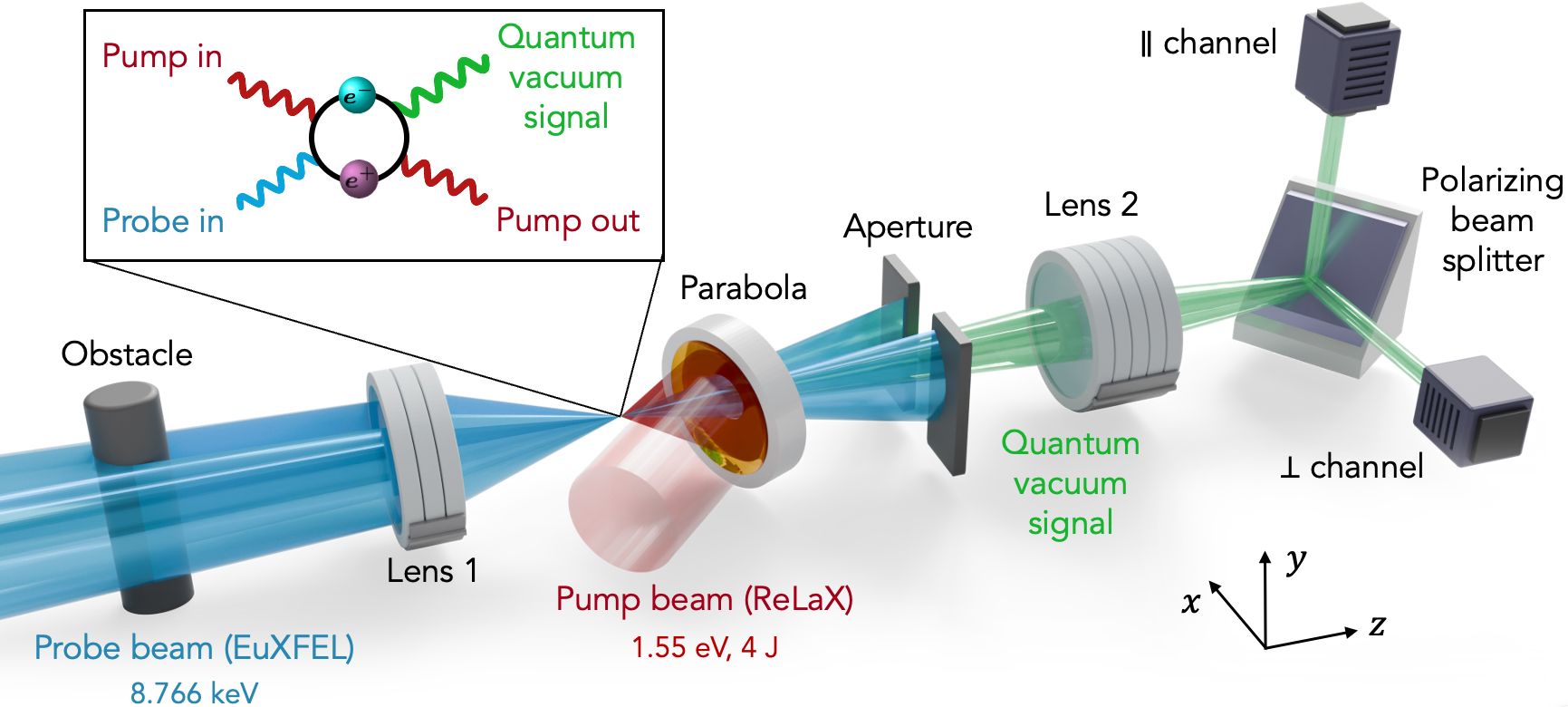}
	\caption{Schematic layout of the dark-field setup implemented in the experiment.
	The near-infrared laser ReLaX is focused with an $f/1.3$ off-axis parabolic mirror featuring a central hole to allow for the collision with the EuXFEL beam in a counter-propagating geometry.
	Its strong electromagnetic fields effectively polarize the quantum vacuum, the nonlinear response of which is probed by the EuXFEL beam.
	Prior to focusing with Lens 1, the incident convergent EuXFEL beam is modified by an obstacle creating a dark-field region in its transverse profile.
	This results in a focus profile with a strong central peak and secondary maxima.
	Outside the focal region, the shadow becomes visible once again. A matched aperture placed at the image plane of the obstacle blocks the outgoing EuXFEL beam.
	The beam waists are chosen such that the induced quantum vacuum signal propagates into a cone overlapping with obstacle shadow and thus passes through the aperture. After collimation with Lens 2,
	a polarizing beam-splitter decomposes the transmitted component into $\parallel$ and $\perp$ polarization channels that are registered with separate detectors.}
	\label{fig:1}
\end{figure*}

This letter reports on the best bound on light-by-light scattering in the low-energy regime ever achieved experimentally  in an all-optical experiment so far. 
To this end, pulses of the near-infrared high-intensity Relativistic Laser at XFEL (ReLaX)~\cite{LasoGarcia:2021} were collided with brilliant x-ray pulses generated by the European X-Ray Free-Electron Laser (EuXFEL) in a counter-propagating pump-probe geometry to detect potential quantum vacuum signals at the x-ray probe energy. Both EuXFEL and ReLaX pulses were linearly polarized at a relative angle of $45^\circ$.
In this case, the relevant contribution to the right-hand side of Eq.~\eqref{eq:EoM} scales linearly with the probe and quadratically with the pump field (see the inset in Fig.~\ref{fig:1}), and consists of components polarized parallel ($\parallel$) and perpendicular ($\perp$) to the incident x-ray probe. The associated signal photon numbers $N_\parallel$ and $N_\perp$ directly quantify this nonlinear quantum vacuum response \cite{Karbstein:2015xra}.

These signal photon numbers $N_{\parallel,\perp}$ are proportional to $(a\pm b)^2$,  yielding the ratio $N_\perp/N_\parallel\simeq(a-b)^2/(a+b)^2$ \cite{Karbstein:2022uwf}.  Consequently, measuring both $N_\parallel$ and $N_\perp$ allows for the extraction  of the individual low-energy coupling constants $a$ and $b$. 
The polarization-flipped component $N_\perp$ directly quantifies vacuum birefringence \cite{Toll:1952rq,Baier:1967zza,Aleksandrov:1985,Heinzl:2006xc,DiPiazza:2006pr}. While differing by polarization-dependent factors, both signal components scale with  the QED light-by-light scattering cross section, $N_{\parallel,\perp} \sim \sigma_{\gamma\gamma}^{\rm QED}$ \cite{Karplus:1950zza,Karplus:1950zz,DeTollis:1964una} (see also \cite{Heinzl:2024cia}). 
The required signal-to-background sensitivity was achieved by implementing a recently proposed \cite{Karbstein:2020gzg,Karbstein:2022uwf} and successfully benchmarked \cite{Smid:2025wmj} dark-field detection concept in combination with a polarization-selective beam splitter for the EuXFEL beam.

Moreover, it has been shown by calculations \cite{Karbstein:2022uwf} that with improved EuXFEL and ReLaX parameters, not only will the QED predictions be directly probed for the first time, but precision measurements of the low-energy coefficients $a$ and $b$ will also become feasible.

\paragraph{Experimental Setup}

The experiment is designed to constrain and ultimately probe the nonlinear response of the quantum vacuum. The vacuum is effectively polarized by the field of a tightly focused, high-intensity laser (the pump) and probed with high-brilliance x-ray pulses \cite{Ahmadiniaz:2024xob} in a counter-propagating geometry predicted to maximize the signal yield.

\begin{figure*}
	\centering
	\includegraphics[width=1\linewidth]{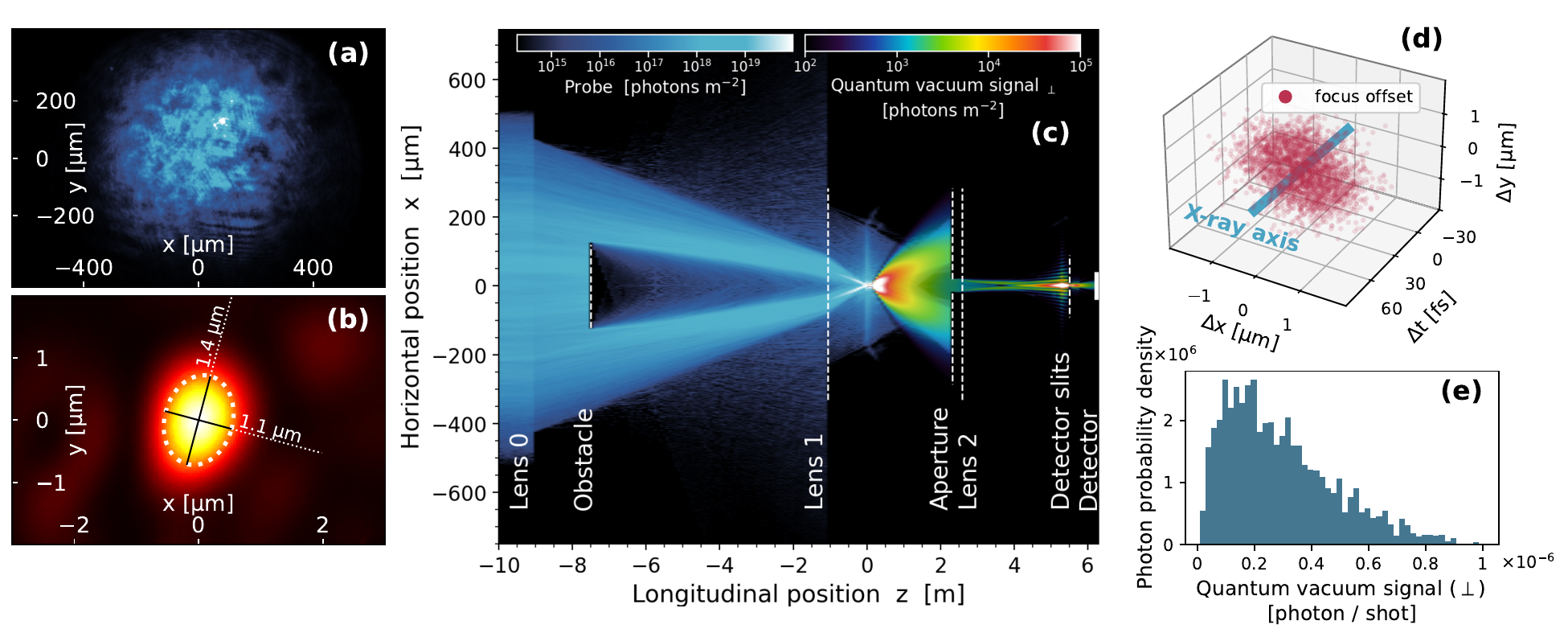}
	\caption{Modeling of the experiment in the simulation tool VIBE \cite{Matheron:2025owt}. The measured EuXFEL intensity profile of a typical shot, prior to the obstacle and ReLaX focus intensity profile used as simulation inputs are shown in (a) and (b).
	Panel (c) shows a two-dimensional top-view cut through a VIBE simulation of the evolution of the EuXFEL probe.  After collision with the ReLaX focus at $z=0$, the induced quantum vacuum signal is visible also on a separate color scale. The simulations account for the details of the setup, including lens imperfections, diffraction and propagation losses. When repeating this simulation for the (d) different focus misalignments between probe and pump (in transverse directions and time) measured on each  shot in the experiment, one can extract (e) the probability distribution of the simulated ($\perp$ polarized) quantum vacuum signal at the detector plane as a function of its signal strength.}
	\label{fig:3}
\end{figure*}

The experiment was performed at the High Energy Density (HED) scientific instrument of the EuXFEL, using the high-intensity ReLaX laser (wavelength \SI{800}{nm}, FWHM pulse duration \SI{23}{fs}, energy at the interaction point \SI{4}{J}) provided by the Helmholtz International Beamline for Extreme Fields (HIBEF) as the pump. Operating the probe at \SI{10}{Hz} and the pump at \SI{0.5}{Hz} yields 19 pump-off reference shots for every pump-on shot under identical beamline conditions. Focusing the pump with an $f/1.3$ off-axis parabolic mirror (OAP) featuring a central hole yielded an elliptical spot with an FWHM minor (major) axis of \SI{1.1}{\micro\meter} (\SI{1.4}{\micro\meter}) (see Fig.~\ref{fig:3}(b) below) corresponding to a peak intensity of \SI{7.7e21}{W/cm^2}.
The EuXFEL delivered seeded pulses at a photon energy of \SI{8.766}{keV} with a typical pulse energy of \SI{150}{\micro\joule} at the interaction point ($N=1.1\times 10^{11}$ photons per pulse) and an FWHM pulse duration of \SI{30}{fs}.

The experimental challenge arises not only from the smallness of the effect but also because the quantum vacuum signal is predominantly scattered in the forward cone of the probe beam. Even under idealized conditions, assuming perfect Gaussian beams and neglecting losses, the maximum polarization-flipped signal per collision for the stated laser parameters is predicted to be on the order of $N_\perp/N\approx10^{-13}$ \cite{Mosman:2021vua}. While this ratio is improved by a factor of $(a+b)^2/(a-b)^2\simeq13.4$ for the $\parallel$ polarized signal component, the latter is even harder to detect because its polarization coincides with that of the probe beam.

A dark-field detection concept, described in detail in~\cite{Karbstein:2022uwf,Smid:2025wmj}, was implemented to separate the  signal from the unscattered x-ray probe (see Fig.~\ref{fig:1}).
Its key idea is to spatially separate the bulk of the induced quantum vacuum signal from the outgoing probe beam and to analyze it at a substantially improved signal-to-background-ratio with a polarization-sensitive detection system.
To this end, the transverse profile of the EuXFEL beam was modified by an obstacle (a vertical polished tungsten wire with a diameter of \SI{250}{\micro\meter}) placed in its beam path prior to focusing with Lens~1, thereby creating a dark-field region in the beam. 
In the focus, the shadow disappears; instead, it is encoded within the side lobes that accompanies the central main peak of \SI{0.9}{\micro m} FWHM width in the direction perpendicular to the obstacle \cite{Chalupsky:2023}.
Interaction with the counter-propagating, tightly focused ReLaX pulse induces a quantum vacuum signal. The transverse source profile of this signal is effectively determined by the convolution of the probe's transverse field profile and the pump's transverse intensity profile.
Proper tuning of the experimental parameters shapes the source profile, inducing a pronounced on-axis signal component during free propagation after the interaction, while the shadow re-emerges for the outgoing probe beam~\cite{Matheron:2026rds}. To isolate the  signal scattered into the dark-field solid angle, an aperture of $\SI{40}{\micro\meter}$ is placed into the image plane of the obstacle. This blocks the probe beam and transmits the quantum vacuum signal.
Lens~2 transports the transmitted photons onto a germanium polarizing beam-splitter \cite{Smid:2025wmj}, which decomposes them into $\parallel$ and $\perp$ polarization channels and further increases the probe beam suppression in the $\perp$ polarization channel.
Both channels are recorded using photon-counting Jungfrau detectors with near-unity quantum efficiency. A detailed description of the experimental setup, characterization, and analysis methods is provided in the accompanying Ref.~\cite{Matheron:2026rds}.

Alignment of the pump and probe foci was established using a $\mathrm{TiO}_2$ screen at the interaction point, which measured their relative spatial position and timing via its optical response. Because this screen must be removed during data acquisition, its measurements were correlated prior to the run with two permanent, non-invasive proxy diagnostics: an equivalent plane camera tracking the transverse misalignment ($\Delta x$, $\Delta y$) and an upstream arrival-time monitor tracking the temporal misalignment $\Delta t$. Together, these diagnostics monitored the misalignment of every shot during the search for the quantum vacuum signal (see Fig.~\ref{fig:3}(d) below). Experimentally, a \SI{0.59}{\micro m} root-mean-square (RMS) focal pointing jitter and a RMS timing jitter of \SI{13.8}{fs} were achieved.

\paragraph{Measurements}

\begin{figure}[t]
	\centering
	\includegraphics[width=1\linewidth]{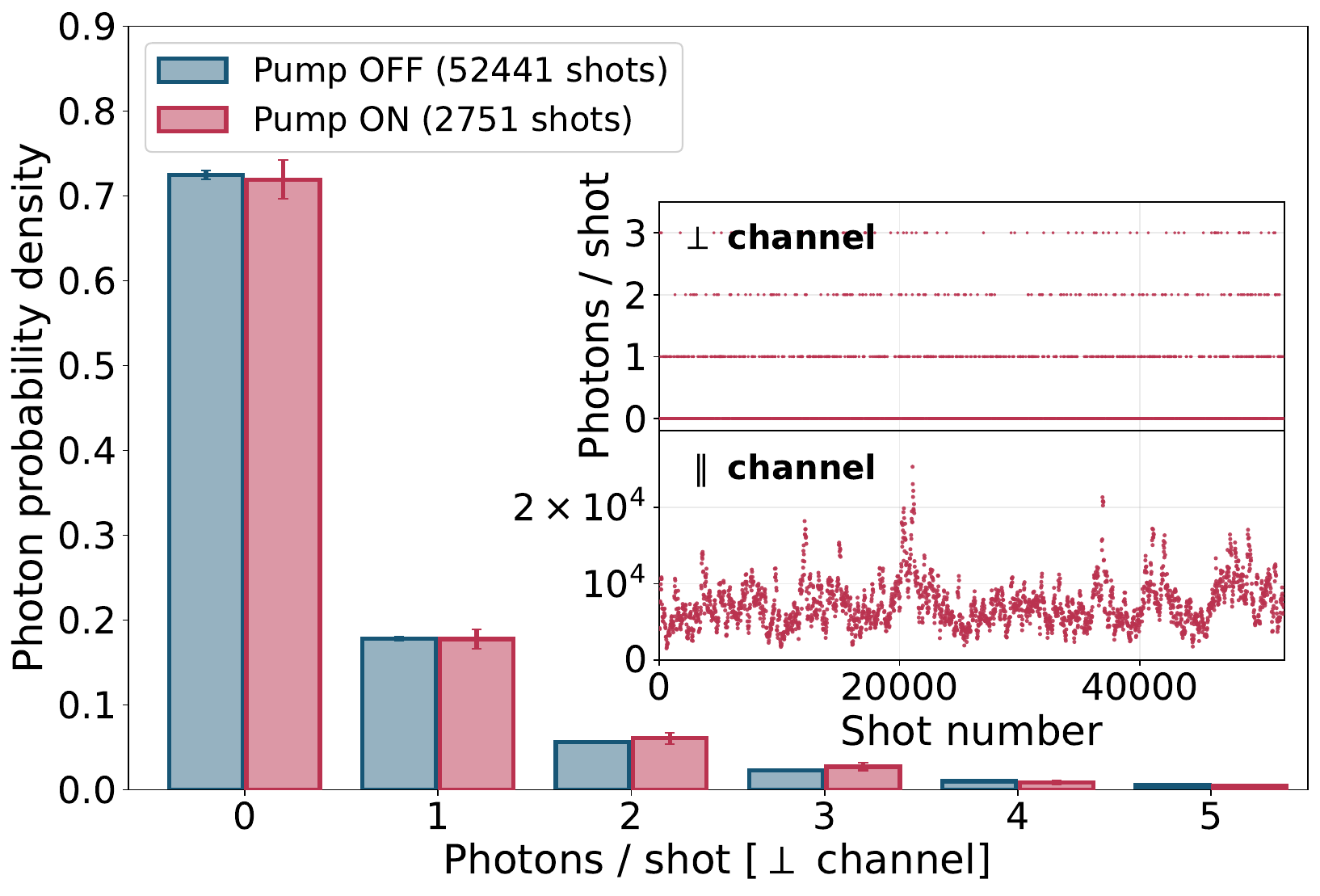}
	\caption{Detected x-ray photons scattered into the solid angle of the dark field. The main panel shows a histogram of the photons detected in the $\perp$ channel for shots without (blue) and with (red) the counter-propagating high-intensity pump, accumulated over $n^{\rm on} = 2751$ pump-on and $52441$ pump-off events; error bars denote $\pm 1\sigma$ statistical uncertainty. The inset shows the photon counts in the $\perp$ and $\parallel$ channels as a function of shot number for pump-on shots; the counts for pump-off shots follow an analogous pattern.}
	\label{fig:2}
\end{figure}

Figure~\ref{fig:2} presents the photon counts recorded in the two polarization channels. Each shot yields several thousand photons in the $\parallel$ channel, with a mean of \num{7264} (\num{7193}) for the pump-on (pump-off) shots and a RMS fluctuation of \num{6724} dominated by intrinsic FEL spectrum, pointing and intensity fluctuations. In contrast, the $\perp$ channel registers much fewer photons per shot, with a mean value of \num{0.4649} (\num{0.4636}) and an RMS fluctuation of \num{0.9570}. A photon count ratio of $6.4\times 10^{-5}$ is measured between the $\perp$ and $\parallel$ channels~\cite{Matheron:2026rds}. Combined with a dark-field suppression of $6.5 \times 10^{-8}$ from the interaction point, this yields a total residual probe transmission in the $\perp$ channel of $4.2 \times 10^{-12}$. The photon distributions for pump-on and pump-off shots are statistically indistinguishable, indicating that no quantum vacuum signal is observed at the current sensitivity level. 

Extracting a quantitative bound from this null observation can be done in the $\perp$-polarization channel where the sensitivity is maximum, and forms the focus of the remainder of this Letter.
The first step is to perform a precise background subtraction. Each pump-on shot $n$ yields a photon count $N_{n}^{\mathrm{on}}$ containing both quantum vacuum signal and background. Each one of these pump-on shot is surrounded by 19 contemporary pump-off shots containing only the background. We estimate the background of the shot $n$ by averaging the 8 nearest-neighbor pump-off shots :
\begin{equation}
\bar N_n^{\mathrm{off}} = \frac{1}{8} \sum_{\substack{j = n-4 \\ j \neq n}}^{n+4} N^{\mathrm{off}}_{j}\,.
\label{eq:local_bkg}
\end{equation}
This local averaging window balances shot-to-shot fluctuation suppression against the XFEL stability timescale, beyond which subsequent pump-off shots decorrelate from the pump-on background. The background-subtracted photon excess of the $n$th pump-on shot is then defined as $N_n^{\rm exp} = N^{\mathrm{on}}_n - \bar N_n^{\mathrm{off}}$.

Averaged over all pump-on shots, the excess in the $\perp$ channel is $-0.0024$~photons. This result is consistent with zero to within the statistical uncertainty $\sigma_{N^{\text{exp}}}/\sqrt{n^{\rm on}} \approx 0.019$~photons, where $\sigma_{N^{\text{exp}}} = 0.994$~photons is the RMS spread of $N_n^{\text{exp}}$. The photon counts in the $\perp$ channel yield a Fano factor~\cite{Cowan:1998ji} of $F=1.97$, indicating super-Poissonian statistics.

\paragraph{Simulations}

To determine the quantum vacuum signal predicted by QED, end-to-end numerical simulations of the experiment were performed using the vacuum emission module VIBE \cite{Matheron:2025owt} within the diffractive beam propagation toolbox LightPipes \cite{LightPipes:1997,LightPipes:2.1.5}.
VIBE tracks the propagation of both the probe and the induced quantum vacuum signal through the entire setup. The simulations are initialized with representative, measured inputs: the transverse intensity profile of the EuXFEL entering the setup [Fig.~\ref{fig:3}(a)], the ReLaX focus profile [Fig.~\ref{fig:3}(b)], and lens thickness-defect maps obtained using x-ray speckle vector tracking~\cite{Berujon:gb5097} during a beamtime at the European Synchrotron Radiation Facility (ESRF) BM05 beamline.
Beyond evaluating the induced quantum vacuum signal under realistic assumptions, the simulations track modifications to both the probe and signal fields caused by individual optical components. Accounting for real-world details, such as scattering from lens impurities, yields a background prediction within the correct order of magnitude of the experimental data.
Figure~\ref{fig:3}(c) shows a two-dimensional spatial cut of a VIBE simulation, tracing the evolution of the EuXFEL beam and the quantum vacuum signal induced during the collision with ReLaX at $z=0$.
In practice, shot-to-shot beam jitter introduces a unique spatio-temporal misalignment between the pump and probe foci for each individual collision.
Repeating the simulation with the experimentally measured shot-by-shot misalignments ($\Delta x, \Delta y, \Delta t$) shown in Fig.~\ref{fig:3}(d) yields the expected QED signal photon number $N^{\mathrm{QED}}_n$, corresponding to each background-substracted experimental count $N_n^{\rm{exp}}$ under the same spatiotemporal misalignment conditions. The probability distribution of $N^{\mathrm{QED}}_n$ for the $\perp$ channel is depicted in Fig.~\ref{fig:3}(e). An integration over the full dataset results in a total expected signal of $N^{\rm{QED}}=7.99\times10^{-4}$ photons in the $\perp$ channel, which translates to a shot-averaged signal of $2.90\times10^{-7}$ photons.

\paragraph{Results}

The analysis centers on the ratio $R = N^{\text{exp}}/N^{\text{QED}}$, where $N^{\text{exp}}$ and $N^{\text{QED}}$ denote the background-subtracted experimental photon excess and the simulated QED signal, respectively. By construction, $R=1$ corresponds to the standard QED prediction. The linear estimator \cite{Cowan:1998ji} employed to determine $R$ and its variance are given by
\begin{equation}
	\hat{R} = \frac{\sum_n N^{\text{QED}}_n N^{\text{exp}}_n}{\sum_n \bigl(N^{\text{QED}}_n\bigr)^2}
	\ \ \text{and} \ \
	\sigma_{\hat{R}} = \frac{\sigma_{N^{\text{exp}}}}{\sqrt{\sum_n \bigl(N^{\text{QED}}_n\bigr)^2}}\,,
	\label{eq:R+sigma}
\end{equation}
where weighting with the simulated counts $N_n^{\mathrm{QED}}$ implements the standard \emph{matched-filter} technique~\cite{Turin:1960}.
This estimator implements an implicit shot selection based on the spatio-temporal alignment of the colliding pump and probe laser pulses, thereby rendering any ad hoc hard thresholding unnecessary. A shot with perfect alignment contributes maximally, whereas poorly aligned shots carrying noise but negligible signal are automatically suppressed.
Applying Eq.~\eqref{eq:R+sigma} to the datasets in Figs.~\ref{fig:2} and \ref{fig:3}(e) yields $\hat{R} = -1.9\times 10^4$ with $\sigma_{\hat{R}} = 5.45\times 10^4$, consistent with a null measurement.
The associated unbounded one-sided upper limit \cite{Cowan:1998ji,ParticleDataGroup:2024cfk} at the 95\% confidence level ($Z = 1.645$), defined as $R_{\mathrm{up}} = \hat{R} + Z\,\sigma_{\hat{R}}$, evaluates to $R_{\mathrm{up}} = 7\times 10^4 $.
This establishes the 95\% confidence level upper bound
\begin{equation}
	\frac{N^{\rm exp}}{N^{\rm QED}}\, \leq\,7\times 10^4 ,
	\label{eq:bound}
\end{equation}
which constitutes the main result of this Letter.
As a simplistic consistency check, the parameters in Eq.~(6) of Ref.~\cite{Smid:2025wmj} (originally derived by Cowan et al. in Ref.~\cite{Cowan:2010js}) are mapped to the variables of this paper by identifying $\#=Z$, $n=n^{\rm on}$, $N_{\rm bgr}=\sigma_{N^{\rm exp}}$, and $N_{\rm sig}=N^{\rm exp}$. Solving the resulting expression for $N^{\rm exp}$ and dividing by the shot-averaged expected QED signal specified above yields an upper bound on the order of $\sim10^5$, which is fully compatible with the bound derived in Eq.~\eqref{eq:bound}.

\begin{figure}
	\centering
	  \begin{overpic}[width=1\linewidth]{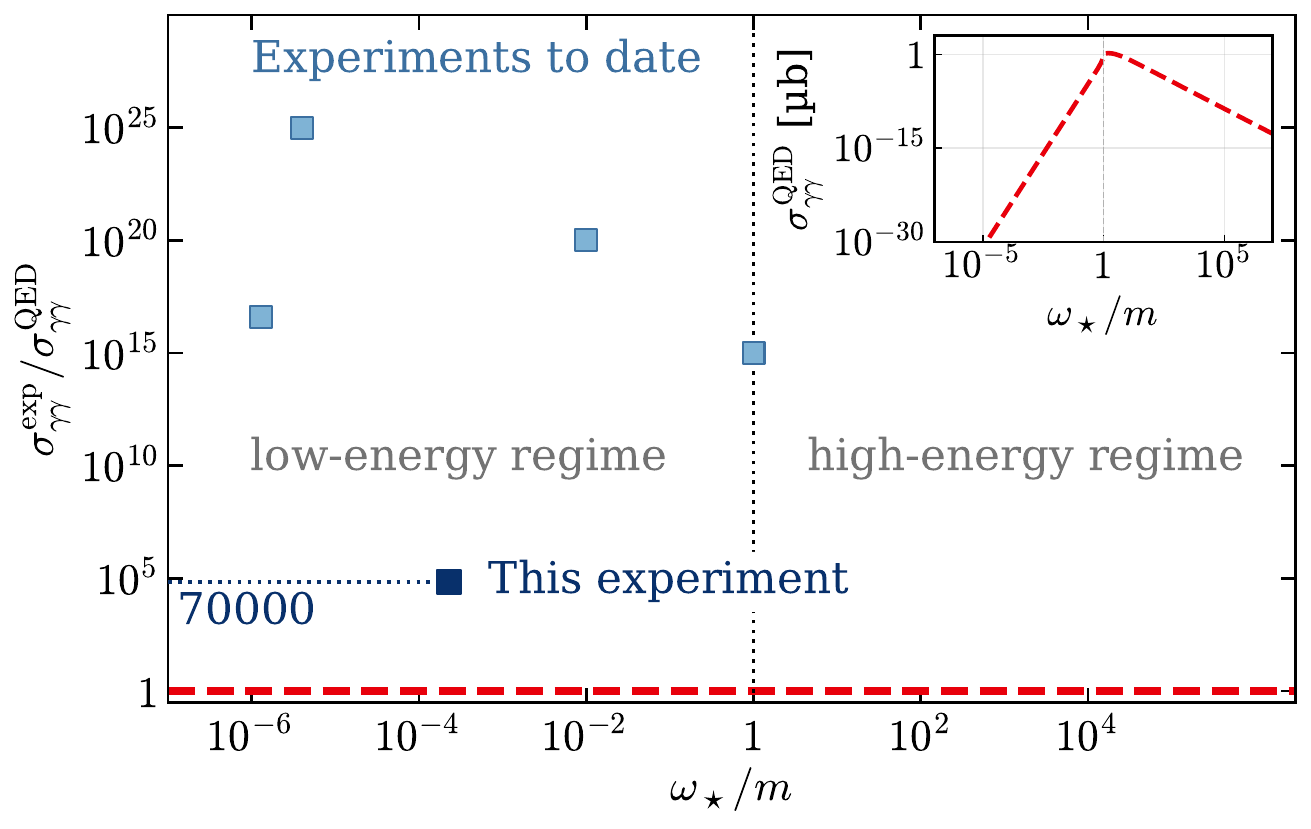}
    \put(22, 38){\color{blue}\cite{Moulin:1996vv}}
    \put(25, 52.5){\color{blue}\cite{Bernard:2000ovj}}
	\put(46, 43.8){\color{blue}\cite{Yamaji:2016xws}}
	\put(59, 35.2){\color{blue}\cite{Watt:2024brh}}
    \end{overpic}
	\caption{Upper bounds (at a $95\%$ confidence level) on the light-by-light scattering cross section, $\sigma_{\gamma\gamma}^{\rm exp}$, inferred from light-beam collision experiments as a function of the center-of-mass photon energy $\omega_\star$. The main panel displays the ratio of $\sigma_{\gamma\gamma}^{\rm exp}$ to the QED prediction $\sigma_{\gamma\gamma}^{\rm QED}$, the behavior of which as a function of $\omega_\star$ is shown in the inset. It is evident that this experiment ($\omega_\star=116\,{\rm eV}$) provides the best bound in the low-energy regime.}
	\label{fig:4}
\end{figure}

The limit derived in Eq.~(\ref{eq:bound}) improves upon previous experimental limits on direct light-by-light scattering. Assuming QED holds, the coherently scattered signal implies a bound on the incoherent scattering process. Using the identity $\sigma_{\gamma\gamma}^{\rm exp}/\sigma_{\gamma\gamma}^{\rm QED}=N^{\rm exp}/N^{\rm QED}$~\cite{Bernard:2000ovj}, we arrive at a new $95\%$ confidence level bound on the cross section: $\sigma_{\gamma\gamma}^{\rm exp}/\sigma_{\gamma\gamma}^{\rm QED}\leq 7\times10^4$  at the center-of-mass photon energy $\omega_\star=\SI{116}{eV}$, given by the square root of the product of the pump and probe photon energies.
Equivalently, a signal exceeding $R_{\mathrm{up}} \sum_n N^{\mathrm{QED}}_n \simeq 56$ photons over the \num{2751} pump-on shots would have indicated a significant excess.
A comparison with established limits in Fig.~\ref{fig:4} highlights that this is by far ($\sim 10^{10}$) the most stringent bound on direct low-energy light-by-light scattering achieved to date.
We do not show the high-energy bounds on light-by-light scattering \cite{dEnterria:2013zqi,ATLAS:2017fur,ATLAS:2019azn,CMS:2018erd} in this figure for three reasons: (i) they probe a different parameter space, (ii) probe the effect via virtual photons, and (iii) their interpretation requires accounting for other SM sectors, going beyond a pure QED framework.

\paragraph{Conclusions and perspectives}

In summary, a dark-field detection scheme has been established at the European XFEL to search for vacuum birefringence and direct light-by-light scattering in the low-energy regime. In the experiment, the x-ray probe was routinely collided with the high-intensity near-infrared ReLaX laser at the HED instrument in a counter-propagating geometry. In combination with a polarizing beam splitter, the dark-field scheme allowed for the simultaneous monitoring of both polarization components of the nonlinear quantum vacuum response with a substantially reduced background noise. Although the baseline QED sensitivity was not reached under current experimental parameters, a new stringent upper bound on light-by-light scattering has been set, significantly advancing beyond previous laboratory limits. The experimental setup constitutes a promising platform towards the first direct measurement of light-by-light scattering signatures in a controlled laboratory environment upon reaching the nominal performance of EuXFEL and ReLaX, and by implementing configurations beyond current facility constraints.

Looking ahead, a substantially improved sensitivity is well within reach using the  same experimental platform at EuXFEL. Operating the ReLaX laser at its published parameters~\cite{LasoGarcia:2021} (\SI{8}{J} energy at the interaction point, \SI{1}{Hz} repetition rate) increases the attainable quantum vacuum signal by a factor of 8 for a fixed measurement interval. Even greater leverage lies in optimizing the background: while the present setup was restricted by facility-specific constraints on obstacle, lens, and detector positions, relaxing these geometry bounds in a successor experiment unlocks alternative dark-field configurations. Exploratory VIBE simulations, benchmarked against the empirical data reported here, demonstrate that these optimized geometries can enhance the dark-field quality by more than three orders of magnitude. Time resolved detectors combined with higher shot rate could also eliminate the effects of short time fluctuations and improve background subtraction by over an additional order of magnitude. Combined, these immediate upgrades are expected to enable the first direct laboratory verification of QED vacuum nonlinearity, including first measurements of photon-photon scattering and vacuum birefringence.

Beyond an initial detection, the dark-field scheme paves the way for new class of precision tests of QED. While in the current background-dominated experiment the ratio $N_\perp/N_\parallel$ merely quantifies the polarization purity, achieving background suppression below the expected signal level in future campaigns will allow this ratio to directly probe the QED prediction $(a-b)^2/(a+b)^2 \simeq0.074$. Together with a measurement of $N_{\perp }$, this approach will enable an independent determination of the dimensionless coefficients $a$ and $b$ in Eq.~\eqref{eq:EoM}, providing a stringent test of their QED predictions in Eq.~\eqref{eq:a,b}.

\acknowledgments
We acknowledge the European XFEL in Schenefeld, Germany, for provision of X-ray free electron
laser beam time at the Scientific Instrument HED (High Energy Density Science) under proposal
number 10692 and would like to thank the staff of European XFEL and DESY for their assistance.
The authors are indebted to the Helmholtz International Beamline for Extreme Fields (HIBEF)
user consortium for the provision of instrumentation and staff that enabled this experiment.
This work has been funded by the Deutsche Forschungsgemeinschaft (DFG, German Research Foundation) under Grants Nos.\ 392856280, 416611371, 416607684, 416702141, and 416708866 within the Research Unit FOR2783/2.

\bibliographystyle{apsrev4-2}

\end{document}